# Analysing Membership Profile Privacy Issues in Online Social Networks

Shafi'i Muhammad Abdulhamid[1], Hassan Abdulazeez[2], Ochoche Abraham[3] and Umar Mohammed[4]

Department of Cyber Security Science, Federal University of Technology, Minna PMB 65, Nigeria[12].

Department of Information Technology, Federal University of Technology, Minna PMB 65, Nigeria[3].

Department of Mathematics and Statistics, Federal University of Technology, Minna PMB 65, Nigeria[4].

E-mail: *shafii.abdulhamid@futminna.edu.ng[1], abochoche@futminna.edu.ng[3] and*

*digitalumar@yahoo.com[4]*

**Abstract**

A social networking site is an on-line service that attracts a society of subscribers and provides such users with a multiplicity of tools for distribution personal data and creating subscribers-generated content directed to a given user's interest and personal life. Operators of online social networks are gradually giving out potentially sensitive information about users and their relationships with advertisers, application developers, and data-mining researchers. Some criminals too uses information gathered through membership profile in social networks to break peoples PINs and passwords. In this paper, we looked at the field structure of membership profiles in ten popular social networking sites. We also analysed how private information can easily be made public in such sites. At the end recommendations and countermeasures were made on how to safe guard subscribers' personal data.

**Keywords:** *Social Networking Sites (SNS), Online Social networks (OSN), Membership Profile, Privacy Issues*

1. Introduction

As defined by Boyd and Ellison (2008), Online Social Networks include web-based services that allow individuals to (1) construct a public or semi-public profile within a bounded system, (2) articulate a list of other users within whom they share a connection, and (3) view and traverse their list of connections and those made by others within their system. Similarly, a recent definition by Hitwise and Experian (2007) states that "Social networking websites are online communities of people who share interests and activities, or who are interested in exploring the interests and activities of others. They typically provide a variety of ways for users to interact, through chat, messaging, email".

Online Social networks are platforms that allow users to distribute personal profiles, connect to each other, upload pictures, blog entries, join groups and look for for friends. Several hundred millions of users join online every year social networks like Hi5, Facebook, Twitter, 9jabook or MySpace. Recently online social networks entail users to place unconditional faith in them, and the inability of the operators to protect users from malicious agents, intruders and masquerados has led to sensitive private information being made public.

Social network operators have databases with such information for millions of users, and their incentives mostly pertaining to growth may not be associated with those of their clients. Users wishing to protect their personal information must have options which do not rely on the all-knowing operators: no matter how incompetent a social network operator is, users should be able to anticipate that their private information is only made public to others when they desire it.

## 2. Related Works

Anderson et al (2009) proposed a design for social networking that protects users' social information from both the network operator and other network users. This design builds a social network out of smart clients and an untrusted central server in a way that eliminates the need for faith in network operators and give users power to manage and protect their privacy. Aside from profiles, Friends, comments, and private messaging, SNSs vary greatly in their features and user base. Many SNS have photo-sharing or video-sharing capabilities; others have built-in blogging and instant messaging technology. There are mobile-specific SNSs (e.g., Dodgeball), but some web-based SNSs also have some degree of mobile interactions (e.g., Facebook, MySpace, and Cyworld). Some SNSs target clients from specific geographical area or linguistic group, even though this does not always indicate the site's population (Boyd and Ellison, 2008).

Data suggest that an increasing volume of cyber crime is being directed to internet users on social networking sites. There have been so many reports of cyber criminals "phishing" for private information on social networking sites through membership profile (Networld, 2011). At risk is not only the personal information of the user, but presumably also that of the user's employer. Unsuspecting users on these sites run the risk of compromising sensitive information, including bank and financial data, highly personal information such as relationship, health and well-being and employment information, and similar sensitive information of family and/or friends (Esecurityplanet, 2011).

Social networking sites allow a potential customer or potential member to simply facilitate a human level relationship with persons within an organization (Carfi and Chastaine, 2002). This enables genuine business relationships to form and puts an authentic human face on the interaction, changing the external perception of an organization from a sterile, faceless behemoth into a collection of individuals who are ready to help. Krishnamurthy and Wills (2010) present taxonomy of ways to study privacy leakage and report on the current status of known leakages. The research found that all mobile social networking sites in the study exhibit some leakage of private information to third parties. Novel concerns include combination of new features unique to mobile access with the leakage in SNS. It is important to recognize the privacy concerns of individuals and organisations. To appreciate the implications of social networks for privacy issues, a number of authors have begun to explore how social networking data can, or can not, be anonymized by means of data perturbation methods. Probabilistic databases can play a motivating role in moving theoretical techniques of privacy-preservation into large scale applications (Adar and R´e, 2007).

## 3. Security Problems in Online Social Networks

The users of an online social network in most cases faithfully expect the system to protect their personal information from made public. Personal information includes both *content* (profile information, post, images, videos, audios, etc.) and *associations* (friendship, group membership, social browsing history, etc.). The system should protect this information from two categories of *attackers:* the network operator and other network users.

The *network operator* of an online social network or the fundamental network facility is an authoritative challenger. In a centralised system, the operator can perform traffic analysis or denial of service attacks without limitations. This is a very serious security problem as the network operators have total access and control over users' personal profile information.

*Other users* of the social network have limited abilities than the network operator: they cannot view all the actions of other users within the network unless they are given access. The most common security threat is simply that they might access content which they are not authorised to access, but most online social networks can prevent them from: accessing social graph data (e.g. friendship links) of other users and editing or deleting another user's information.

## 4. Online Social Networks

The are a number of excellent online social networking sites available to anyone interested in becoming part of the social networking community of the internet. But for the purpose of this research, ten (10) of the online networking sites have been selected in no particular order. These includes:

### 4.1 Facebook (www.facebook.com)

Facebook arguably the most popular social networking sites in the cyberspace. Beginning in September 2005, Facebook expanded to include high school students, professionals inside corporate networks, and, eventually, everyone. Facebook users are unable to make their full profiles public to all users. Another feature that differentiates Facebook is the ability for outside developers to build "Applications" which allow users to personalize their profiles and perform other tasks, such as compare movie preferences and chart travel histories (Facebook, 2011).

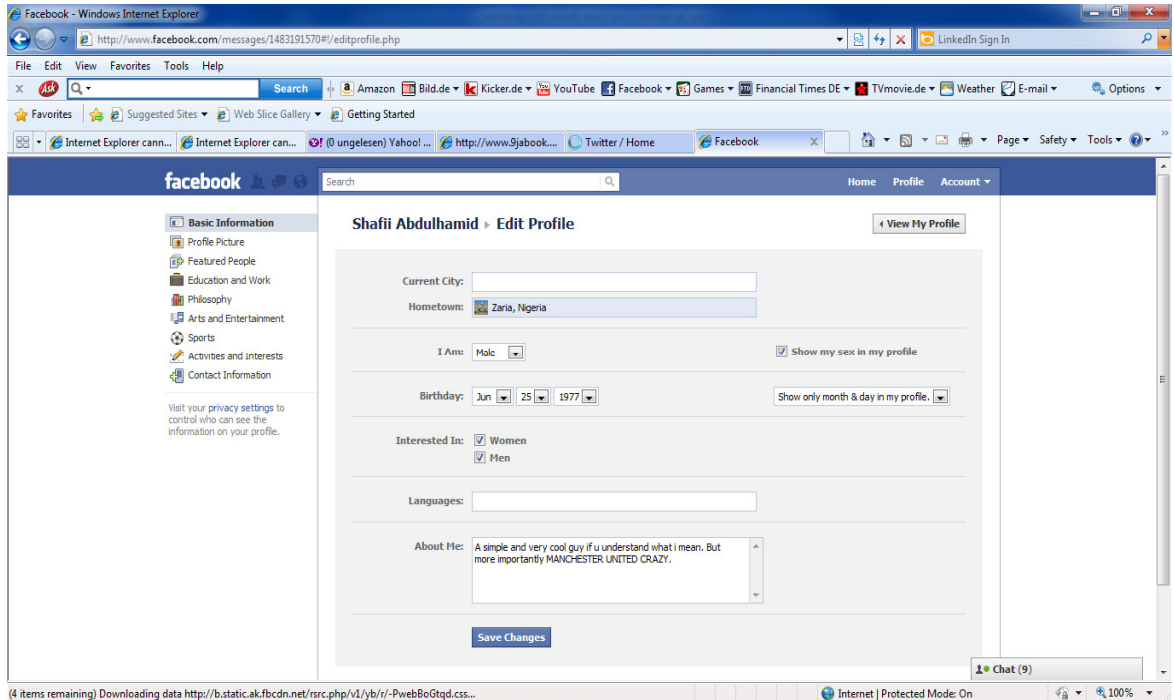

**Figure 1:** *Facebook Interface*

**4.2 Twitter (www.twitter.com)**

Twitter (founded 2006) is a real-time information social network that connects users to the latest information about what the user find interesting. Simply the user find the public streams is most compelling and follow the story. At the heart of Twitter are small bursts of information called *Tweets*. Each *Tweet* is 140 characters long. Connected to each *Tweet* is a rich details pane that provides additional information, deeper context and embedded media. The user can tell story within the *Tweet*, or think of a *Tweet* as the headline, and use the details pane to tell the rest with photos, videos and other media content (Twitter, 2011).

**4.3 Hi5 (www.hi5.com)**

Hi5.com was launched into full operations in the year 2003. It operates on similar ways with Facebook. The user can connect with friends, search for friends and share information with them, these includes personal profile viewing, images, videos and even audios.

**4.4 LinkedIn (www.LinkedIn.com)**

LinkedIn.com was conceived and started right at the living room of co-founder Reid Hoffman in 2002. The site was officially launched on May 5, 2003. At the end of the first month in operation, LinkedIn recorded 4,500 members in the network. It took 494 days to reach the first million members, and now, on average, a new member joins every second of every day, or approximately one million every 12 days. The company is privately held and has a diversified business model with revenues coming from user subscriptions, advertising sales and hiring solutions (LinkedIn, 2011).

**4.5 YouTube (www.youtube.com)**

YouTube.com was first launched in February 2005, it allows millions of users to discover, watch, upload, download and share originally-created videos. The site provides a forum for users to connect, inform, and encourage others across the world and acts as a sharing platform for innovative content creators and advertisers of various sizes (YouTube, 2011).

**4.6 Friendster (www.friendster.com)**

Friendster (founded 2002) is one of the pioneer and leading global social networking site, is focused on helping people stay in touch with friends and discover new things and people. People can easily join with anyone around the globe via *www.friendster.com* or *m.friendster.com* from any Internet-ready mobile device. Friendster was acquired in December 2009 by MOL Global, the parent company of Asia's leading online payment solutions provider MOL AccessPortal Berhad (Friendster, 2011).

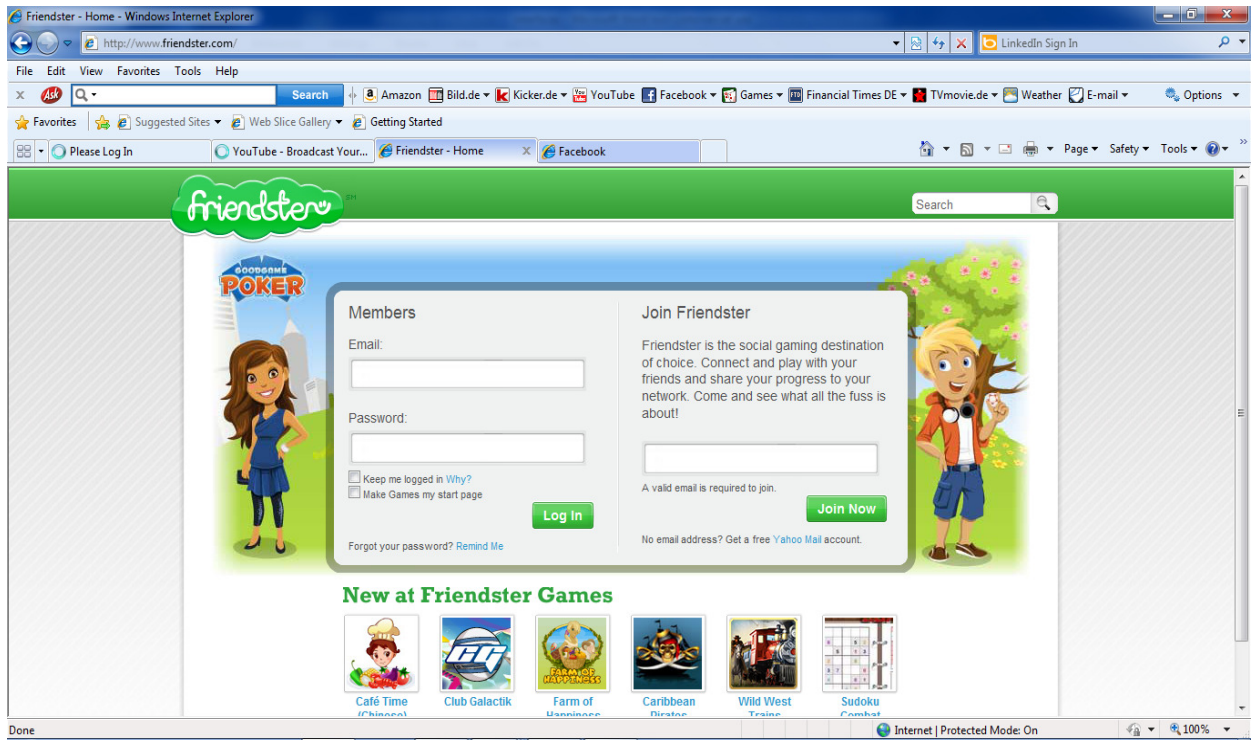

**Figure 2:** *Friendster Interface*

**4.7 Tribe (www.tribe.net)**

Tribe.net was founded by Mark Pincus, Paul Martino and Valerie Syme. It connects you to friends, people who live in your locality or languages, or people who live in your city that have common curiosity. The user can invite friends, search for people with similar interests, and join or create tribes (member-created online groups) dedicated to interests you might have. The more people you're connected to, the enhanced tribe.net will be for the user (Tribe.net, 2011).

**4.8 MySpace (www.myspace.com)**

MySpace was launched in the year 2003. It is described as an online network for friends to meet their friends. It is one of the most popular social networking sites and allows its users to network, to share photos, videos and to create journals, amongst other things. It records 110 million monthly active users around the globe.

**4.9 Bebo (www.bebo.com)**

Bebo is a fashionable social networking site (SNS) which connects you to everyone and everything you like. It is your life on the cyberspace – social experience that assist you to discover what's just about your world and helps the world discover what's just about you. Bebo combines group of people, self-expression and amusement, enabling you to consume, create, discover, curate and share e-content in entirely new ways. Bebo has sites all across the world (Bebo, 2011).

**4.10 9jabook (www.9jabook.com)**

The main motive behind the creation of 9jabook.com is to participate in redeeming the image of Nigeria to the rest of the world. This is a product of the "Rebranding Nigeria campaign". 9jabook.com is also a stylish SNS used to connect with friends and also share news information. It was founded in the year 2009.

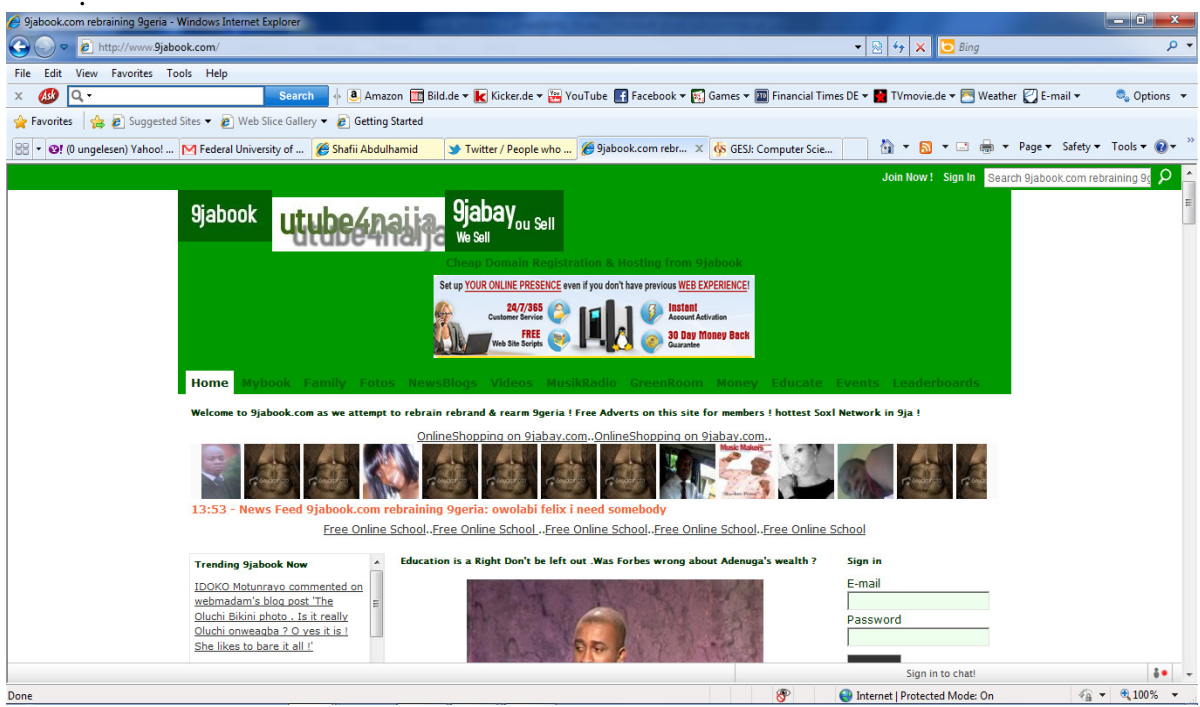

**Figure 3:** *9jabook Interface*

## 5. Methodology

The methodology adapted in this research work was using internet data mining and personal observations. Ten (10) popular social networking sites (SNS) were selected for the research in no particular order. The authors registered with all the ten SNS selected with different user names for a period of three months. We connected to make friends and perform all other operations like a normal user and over the time the observations made were recorded. Also, data mined or obtained directly from the websites were also recorded and tabulated for presentation and analysis. The period under study was 10$^{th}$ January, 2011 to 10$^{th}$ April, 2011.

Table 1 below presents the data obtained from the membership profiles of the ten SNSs under study as categorized according to the fields or features of the profile. In the table below, 1 is used to represent a field that exists in the SNS and a blank or 0 is used to represent a field that did not exist in the SNS. In the last column, a total was calculated from the ten SNSs for each field or feature. The last row of the table, presents the total fields that each site has, the number of fields are not limited to the ones listed below.

Table 2 below presents the data obtained directly from the sites stating the approximate number of registered users in each of the SNS.

## 6. Results and Discussion

The Table 1 below shows the actual data as it was extracted from the ten SNSs under study. The data was used to plot the bar chart in Figure 4 below.

| S/No. | Profile Information | Facebook | Twitter | Hi5 | LinkedIn | YouTube | Friendster | Tribe | MySpace | Bebo | 9jabook | Total |
|---|---|---|---|---|---|---|---|---|---|---|---|---|
| 1 | Photo | 1 | 1 | 1 | 1 | 1 | 1 | 1 | 1 | 1 | 1 | 10 |
| 2 | Professional Detail | 1 | 1 | | 1 | 1 | | 1 | | 1 | 1 | 7 |
| 3 | Gender | 1 | 1 | 1 | 1 | 1 | 1 | 1 | 1 | 1 | 1 | 10 |
| 4 | Age/Date of Birth | 1 | 1 | 1 | 1 | 1 | 1 | 1 | 1 | 1 | 1 | 10 |
| 5 | Sexual Orientation | | | | | 1 | | | 1 | | | 2 |
| 6 | Marital Status | 1 | 1 | 1 | 1 | 1 | 1 | | 1 | 1 | 1 | 9 |
| 7 | Sense of Humor | | | | | 1 | | | | | | 1 |
| 8 | Hobbies/Interest | 1 | 1 | 1 | 1 | 1 | 1 | 1 | 1 | | 1 | 9 |
| 9 | Favorite Music | 1 | | 1 | 1 | 1 | 1 | 1 | 1 | 1 | 1 | 9 |
| 10 | Favorite TV | 1 | 1 | | | 1 | 1 | 1 | 1 | | | 6 |
| 11 | Favorite Books | 1 | 1 | | | 1 | 1 | 1 | 1 | | | 6 |
| 12 | Favorite Food | | | | | 1 | | | | | | 1 |
| 13 | Location | 1 | 1 | 1 | 1 | 1 | 1 | 1 | 1 | 1 | 1 | 10 |
| 14 | Home Town | 1 | 1 | 1 | 1 | 1 | 1 | 1 | 1 | 1 | 1 | 10 |
| 15 | Here for… | 1 | | | 1 | 1 | 1 | 1 | 1 | | | 6 |
| 16 | College/University | 1 | 1 | 1 | 1 | | 1 | | | 1 | 1 | 7 |
| 17 | Clubs & Organizations | 1 | 1 | | | | | 1 | | | 1 | 4 |
| 18 | Languages | 1 | 1 | 1 | 1 | | | 1 | | 1 | 1 | 7 |
| 19 | Religion | 1 | 1 | | 1 | 1 | | | 1 | | 1 | 6 |
| 20 | Smoking | | | | | 1 | 1 | | 1 | | | 3 |
| 21 | Drinking | | | | | 1 | 1 | | 1 | | | 3 |
| 22 | Nationality | 1 | 1 | 1 | 1 | 1 | 1 | 1 | 1 | 1 | | 9 |
| | **Total** | **17** | **15** | **11** | **14** | **19** | **15** | **14** | **16** | **11** | **13** | |

**Table 1:** *Membership Profile Fields Obtained From SNSs*

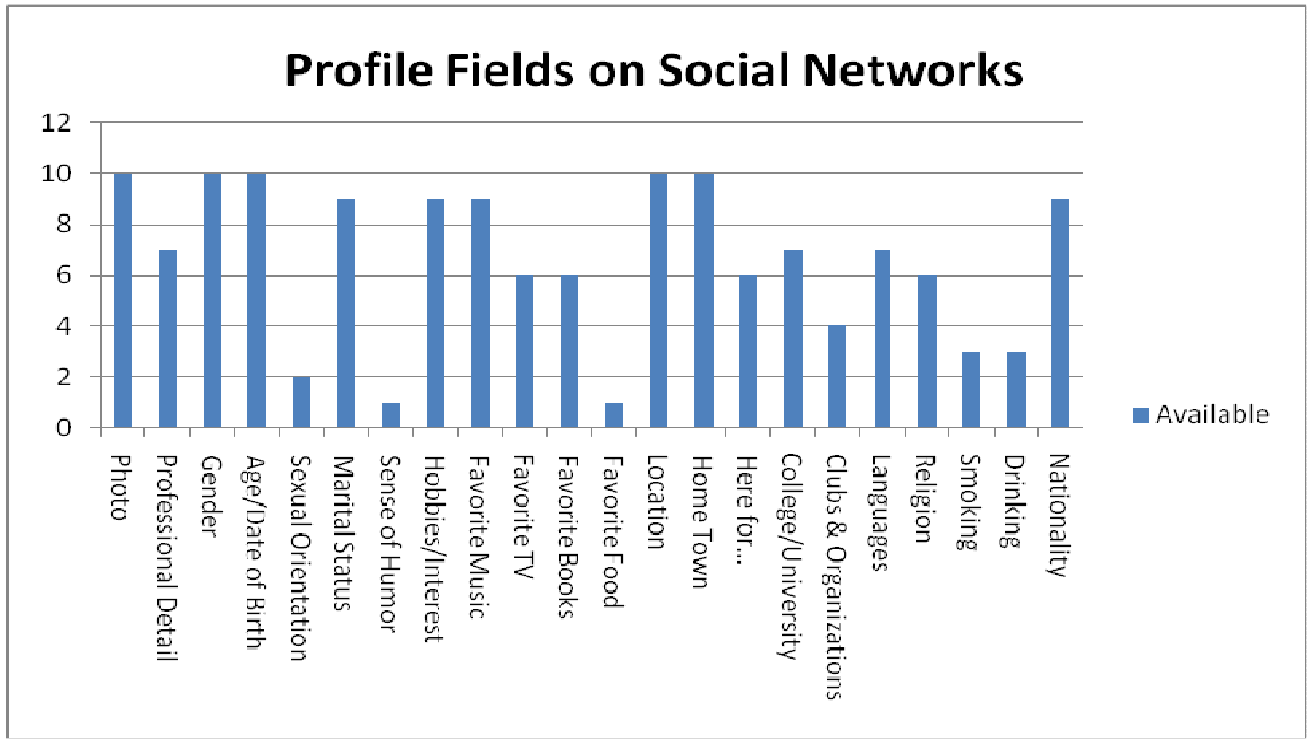

**Figure 4:** *Bar Chart Showing the Membership Profile Fields*

From Table 1 and Figure 4 above, the following interesting findings can be inferred:

- All the ten SNSs demand for photos, gender, age or date of birth, location and town names. This implies that most SNS want the users to disclose their facial identity, sex, date of birth and town. It is a common knowledge that most people use what they can remember easily as their Personal Identification Number (PIN) for credit cards and ATMs and also for e-mail addresses. A combination of date of births is one way to guess PIN and town names is another way to also guess e-mail passwords.

- The fields with second highest number of appearance in most of the SNSs are marital status, favorite music, hobbies and nationality. This information can be very helpful in cracking PINs and e-mail accounts, as some users uses their hobbies and names of favourite music and artists as passwords. The people of certain countries are also associated with some positive and negative things, like Indians with movies and cobras, Arabs with Islam and terrorism, Italians with soccer and mafiarism, Americans with country music, rock and propaganda, Nigerians with big heart and 419 – these types of information can also be very helpful to security agencies during investigations.

- Favorite food, sense of humor and sexual orientation recorded the least number of appearances in the ten SNSs under study. These are little information that reveals more about individual privacy.

- Professional details, College/University attended and religion are information that reveals many things about person's educational level, cultural and academic experiences. It also says a lot about faith and religious inclinations. Some professionals use things that they are more familiar with in their fields of work as PIN and passwords. Example a mathematician may use the value of mathematical constants like $\pi = 3.1428571429$ or a physicist may use formulas like $e = mc2$. This may also be a clue to hackers.

- Profiles on SNSs can be downloaded and stored over time and incrementally by third parties, designing a database of personal data. Information exposed on an SNS can be used for purposes and in contexts poles apart from the ones the profile owner had considered.

Table 2 and Figure 5 below shows the data mined from the ten SNSs under study. The data indicates the approximate number of registered users of each of the SNS as at the period under review 10th January, 2011 to 10th April, 2011. The bar chart shows the pictorial representation of the data.

| S/NO. | Name of SNS | Registered Users |
|---|---|---|
| 1 | Facebook | 500,000,000 |
| 2 | Twitter | 175,000,000 |
| 3 | Hi5 | 200,000,000 |
| 4 | LinkedIn | 90,000,000 |
| 5 | YouTube | 190,000,000 |
| 6 | Friendster | 85,000,000 |
| 7 | Tribe | 23,000,000 |
| 8 | MySpace | 130,000,000 |
| 9 | Bebo | 66,000,000 |
| 10 | 9jabook | 300,000 |

**Table 2:** *Approximate Number of Registered Users*

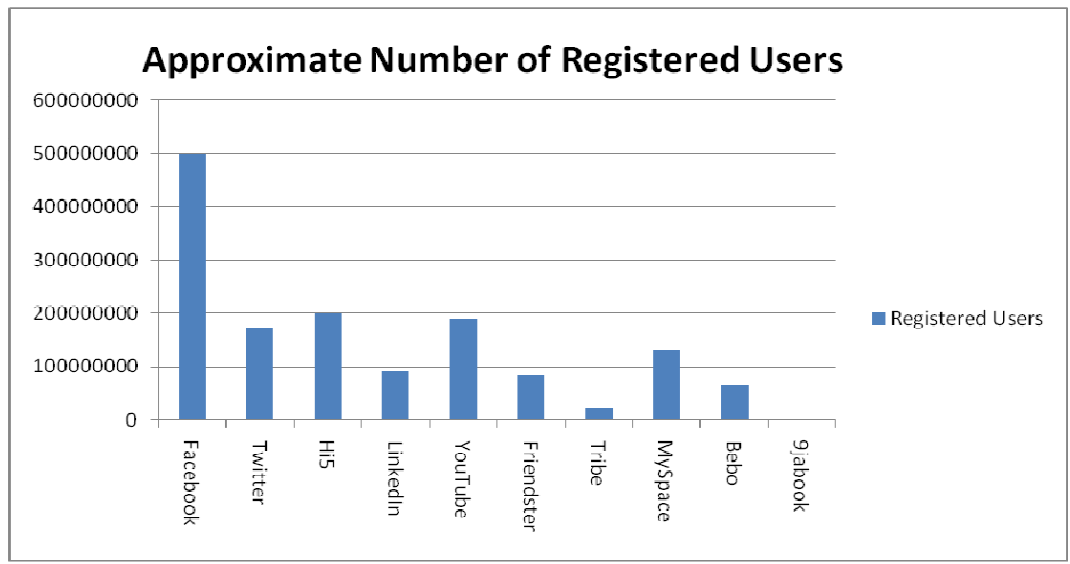

**Figure 5:** *Bar Chart Showing the Approximate Number of Registered Users*

From Table 2 and Figure 5, the following interesting findings can be inferred:

- From the data presented above, the question everybody should be asking is "*which SNS will pose the greatest profile and security threat?*". Although networks with high number of registered users present greater opportunities for socialising, it also presents the greater risk of information leakage. So the answer to this is trivial, as Facebook have the heighest number of users and activity. By implication it means 9jabook may constitute the least number of activity and risk.

- Data harvesters normaly target places where their harvest will yield more results, therefore, SNSs like Facebook, Hi5, Twitter, YouTube and MySpace are their primary targets.

- From the data above, 9jabook and Tribe have a very small number of registered users. Some very smart hackers prefer to target relatively new or unpopular sites, this is because it is easy to discover programing lapses in such site. Users and developers of such sites are generaly more careless and less experience because they deal with less population than the mega SNS like Facebook and Twitter.

### 7. Recommendations and Countermeasures

The first question we want to ask here is "do police force have the right to access content posted to SNS without a warrant?" The legality of this hinges on users' anticipation of privacy and whether or not SNS profiles are considered public or private – this problem will definitely have to go through constitutional amendments in many countries to solve.

How many users of SNSs bother to read carefully their privacy policy? A statement in Facebook privacy policy states "*Removed information may persist in backup copies for a reasonable period of time but will not be generally available to members of Facebook* (Facebook, 2011, b)". This means that deleted information are retrievable by the operators and it remain in their database for as long as they wish. Another privacy policy states that *"Your name, network names, and profile picture thumbnail will be available in search results across the Facebook network and those limited pieces of information may be made available to third party search engines* (Facebook, 2011, b)*"*. With these types of laws, it is advisable that the users should think very well before supplying any type of information to any SNS.

Users of ONS are strongly advice to take advantage to strong privacy and security settings provided by the networks to better protect their profiles and information (example is hide important fields). It is important users should try as much as possible to avoid suspicious third party applications. They should treat everything as public and share information with only people they know – best practice.

## 8. Conclusion

The ever increasing prospects of SNSs can not be over emphasized in this paper and if used securely can enhance data privacy and contributory contents management, even better than portals, blogs and websites. This technology also provides a dangerous powerful tool to people who want to take criminals advantage of unsuspecting users by misusing their membership profile data. False sense of intimacy and masquerading is another problem faced by SNSs and this can be handled through the introduction of facial recognition tools which will improve personal and even physical privacy.

Online social networks are too important to be left in the hand of operators alone. Therefore, Governments have to come in and provide appropriate legislation that will help in protecting users private information. The laws should also protect the operators from having extreme power and control over membership profiles and posts. The Government should also help in creating awareness on the danger of privacy leakage in SNS. This can be done through schools, medias and parent-child training at homes.

In conclusion, authorities will have to play a major role in how secure social networks are, with much greater efforts required to crack down on current cybercriminals and discourage new blood from joining the dark side. These efforts must be implemented at both national and global level to ensure that crimes and criminals cannot be harbored and abetted by rogue nations ignoring global regulation.

**ABOUT THE AUTHORS**

**Shafi'i Muhammad ABDULHAMID** holds M.Sc. degree in Computer Science from Bayero University Kano, Nigeria (2010) and B.Tech. degree in Mathematics/Computer Science from the Federal University of Technology Minna, Nigeria (2004). Presently he is a lecturer at the Department of Cyber Security Science, Federal University of Technology Minna, Niger State, Nigeria.

**Hassan ABDULAZEEZ** holds B.Tech. degree in Mathematics/Computer Science from the Federal University of Technology Minna, Nigeria (2006). Presently he is a lecturer at the Department of Cyber Security Science, Federal University of Technology Minna, Niger State, Nigeria.

**Ochoche ABRAHAM** holds a Ph.D. in Numerical Methods for Ordinary Differential Equations(2010), M.Tech. degree in Mathematics and B.Tech. degree in Mathematics/Computer Science all from the Federal University of Technology Minna, Nigeria. Presently he is the Head of Department of Information Technology, Federal University of Technology Minna, Niger State, Nigeria.

**Umar MOHAMMED** holds M.Tech. degree in Mathematics (2010) and B.Tech. degree in Mathematics/Computer Science all from the Federal University of Technology Minna, Nigeria (2006). Presently he is a lecturer at the Department of Mathematics and Statistics, Federal University of Technology Minna, Niger State, Nigeria.